# Internet Banking System Prototype

Rami Alnaqeib, Hamdan.O.Alanazi, Hamid.A.Jalab, M.A.Zaidan, Ali K.Hmood

**Abstract** – Internet Banking System refers to systems that enable bank customers to access accounts and general information on bank products and services through a personal computer or other intelligent device. Internet banking products and services can include detailed account information for corporate customers as well as account summery and transfer money. Ultimately, the products and services obtained through Internet Banking may mirror products and services offered through other bank delivery channels. In this paper, Internet Banking System Prototype has been proposed in order to illustrate the services which is provided by the Bank online services.

——————————— ◆ ———————————

## 1. INTRODUCTION

System prototyping is the development of incomplete representations of a target system for testing purposes and as a way of understanding the difficulties of development and the scale of the problem. Prototyping is an essential element of an iterative design approach, where designs are created, evaluated, and refined until the desired performance or usability is achieved[1],[2]. Prototypes can range from extremely simple sketches (low-fidelity prototypes) to full systems that contain nearly all the functionality of the final system (high-fidelity prototypes) [3].

## 2. USER INTERFACE DESIGN

User interface design or user interface engineering is the design of computers, appliances, machines, mobile communication devices, software applications, and websites with the focus on the user's experience and interaction [4]. Where traditional graphic design seeks to make the object or application physically attractive, the goal of user interface design is to make the user's interaction as simple and efficient as possible, in terms of accomplishing user goals—what is often called user-centered design. Where good graphic/industrial design is bold and eye catching, good user interface design is to facilitate finishing the task at hand over drawing attention to itself [5]. Graphic design may be utilized to apply a theme or style to the interface without compromising its usability [6]. The design process of an interface must balance the meaning of its visual elements that conform the mental model of operation, and the functionality from a technical engineering perspective, in order to create a system that is both usable and easy to adapt to the changing user needs. In order to provide user-friendly and an intuition interface, several standards will be maintained in the areas of screen navigation, data-entry procedures and button activation [6].

### 2.1 COMMON FIELD REQUIREMENTS

- E-mail Address: Email addresses must contain an @ sign and a period (.). As an example : user@hotmail.com.
- Passwords: Passwords can be 6 to 20 characters in length and consists of alpha characters, numbers and non-alphanumeric characters (#, ^,*, etc.) combined.

### 2.2 COMMON BUTTON FORMATS

Common buttons have the same look and feel, and use standard text:
Add: Adds a new record
Update: Saves changes to an existing record
Delete: Remove a record
Reset: Removes changes to data-entry fields and returns the field values to their prior settings within the record
Submit: Sends a request or initiate an action
Logout: Returns the Administrative User to the Internet Banking log-on screen.

- **Browser**

Internet Explorer 5.5/6.0 or higher (Internet Explorer 6.0 is recommended), Mozilla Firefox, Netscape Communicator, Opera, Safari (for Mac)

## 2.3 LOW-FIDELITY PROTOTYPE

- A low-fidelity prototype is a 'quick and dirty' mockup that is cheap, easily changed, and can be thrown away without complaint.
- During usability tests, a low-fidelity prototype often uses a person as the computer and a pointer as the mouse.
- The goal of such a prototype is to create something as quickly as possible that will elicit user feedback [10].
- Very often, paper and pencil are used to construct this type of prototype, though presentation software (e.g., PowerPoint) may be used [11].

As an early-design tool, a paper low-fidelity prototype is ideal. Many ideas can be viewed and evaluated by the design team in a short period of time, and with very little cost [12]. The basic idea is to have the design team work together, using little more than pen-and-paper, to draw the screens needed for a basic user interface to the product. Because it is done early, quickly, and with no expectation of creating a working version, the team is under much less pressure and generally works together more smoothly [13]. In addition, little to no attachment for the prototype develops in the team, resulting in much less resistance to change. A paper prototype can also be used for a usability test. The major difference with regular usability testing is that a person acts as the computer, changing screens, vocalizing error messages, etc. The low-fidelity prototype of this internet banking system is shown in the appendix [7],[14].

## 2.4 HIGH-FIDELITY PROTOTYPES

- A high-fidelity prototype is much closer to the actual product in look and feel
- The prototype often requires use of a programming language
- The user interacts with a computer, not a person acting as a computer.

High-fidelity prototypes are often written using HTML, JavaScript, VB Script or other programming languages. It is common for the prototype to be written in the same language as the final product, but without the complete functionality or 'clean' code of a final product. This makes the transition from prototype to product quicker and easier [8]. It also increases the risk that the prototype's 'cludge' code will inadvertently be used in the final product. In this internet banking system high-fidelity prototype, the web portal was developed by using ASP application with Macromedia Dreamweaver and the database was developed by using MySql. The web portal was then uploaded to the server and it can be accessed through the address below for testing and evaluation. The system user interface prototype can be shown in figures below [9]. Username: user, Password: user

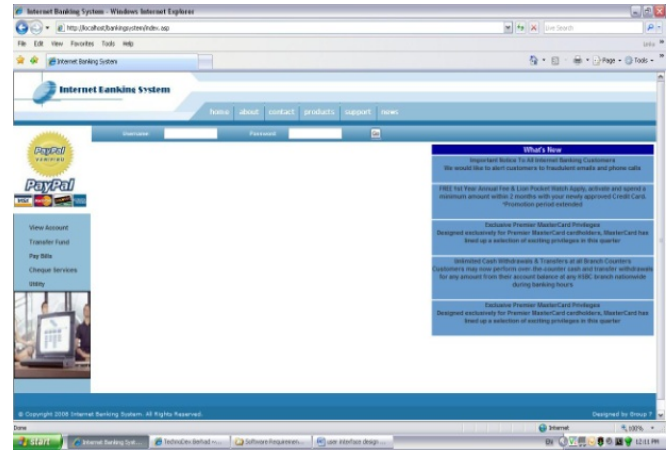

Fig 1. User Interface – Login Page

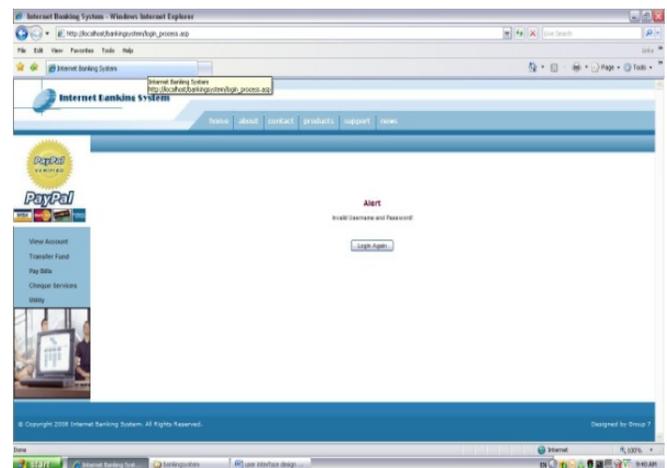

Fig 2. User Interface – Invalid User Page If Username and Password are Incorrect

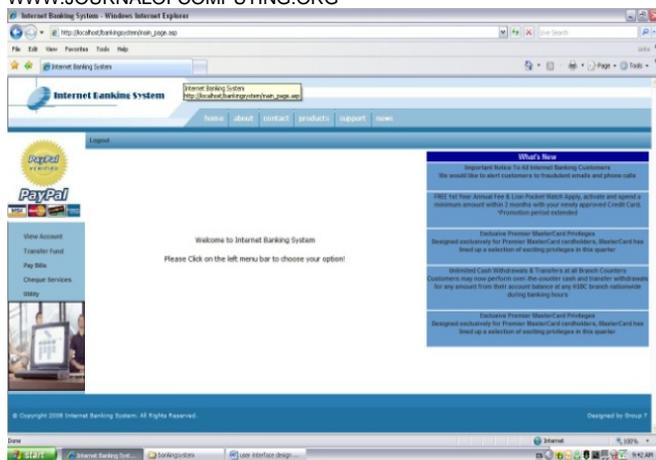

Figure 3: User Interface – User Main Page if Username and Password is Correct

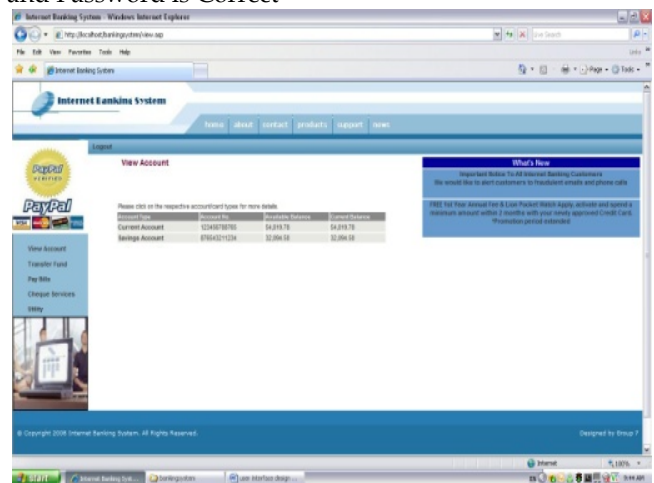

Fig 4. View Account – View Account Main Menu Shows Current Account and Savings Account Balances

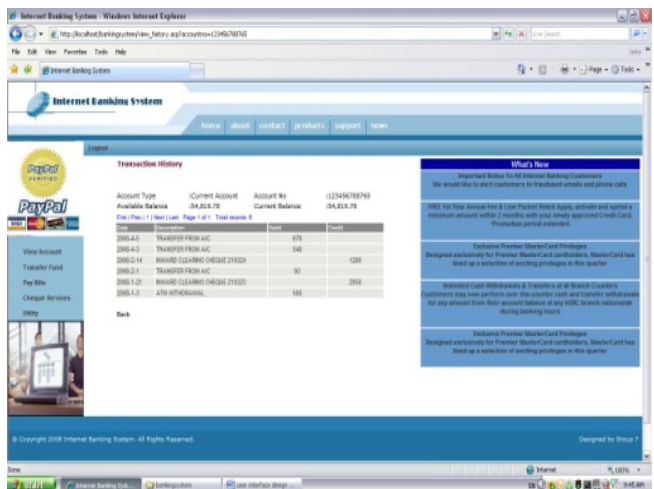

Fig 5. View Account – Transaction History for Current Account

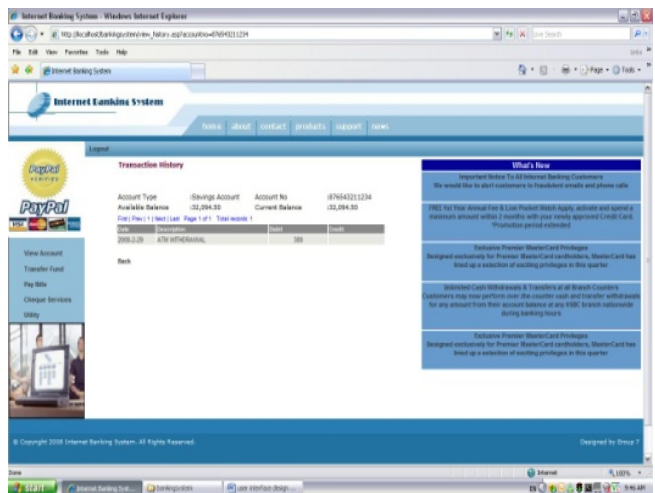

Fig 6. View Account – Transaction History for Savings Account

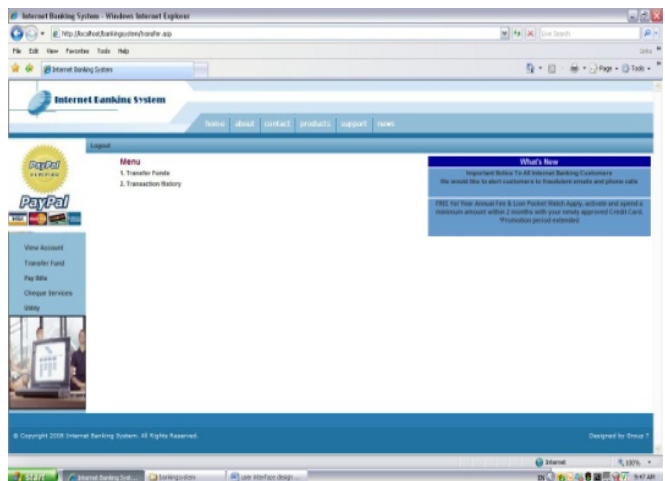

Fig 7. Transfer Funds – Transfer Funds Main Menu

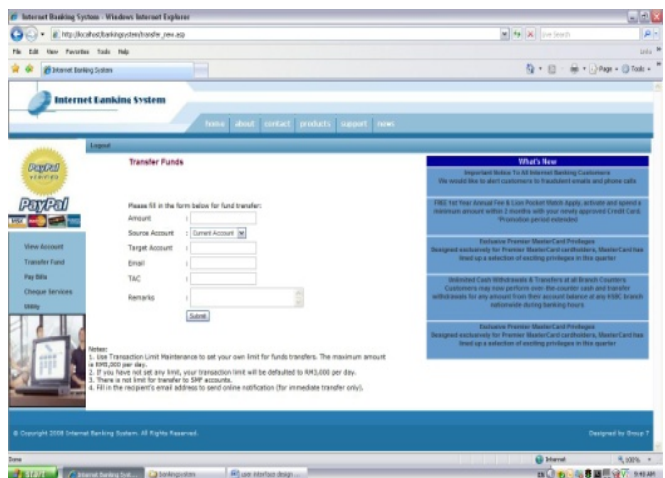

Fig 8. Transfer Funds – Transfer Funds Form

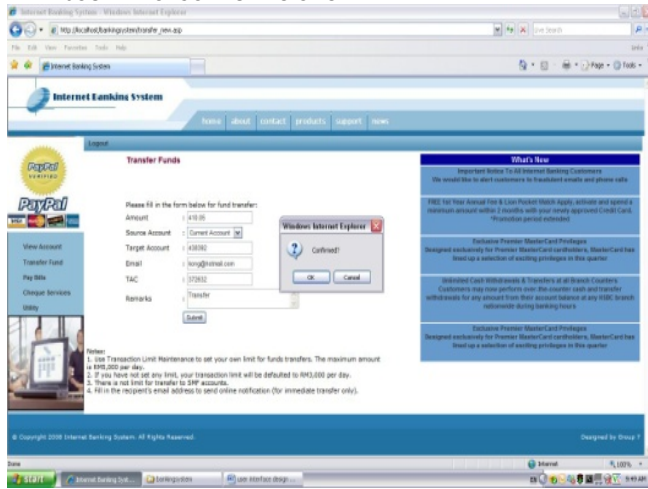

Fig 9. Transfer Funds – Transfer Funds Confirmation Pop Up Message

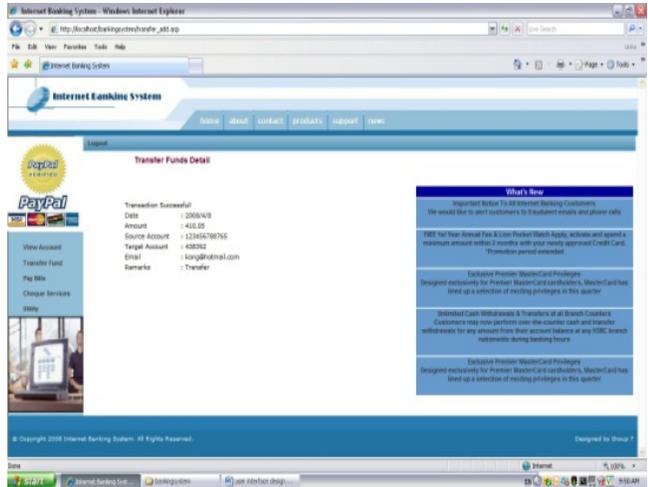

Fig 10. Transfer Funds – Transaction Successful

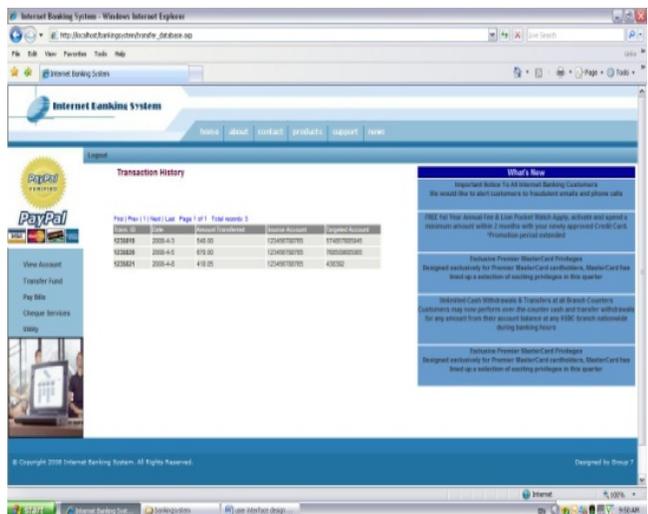

Fig 11. Transfer Funds – Transaction History for Transfer Funds

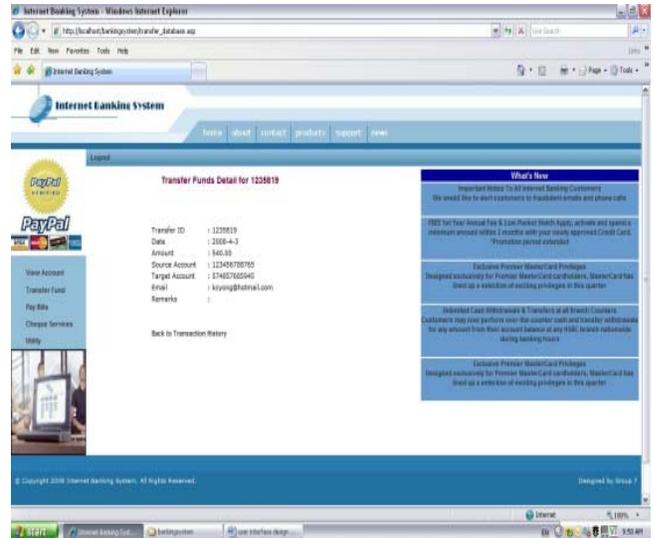

Fig 12. Transfer Funds – Transfer Funds Detail

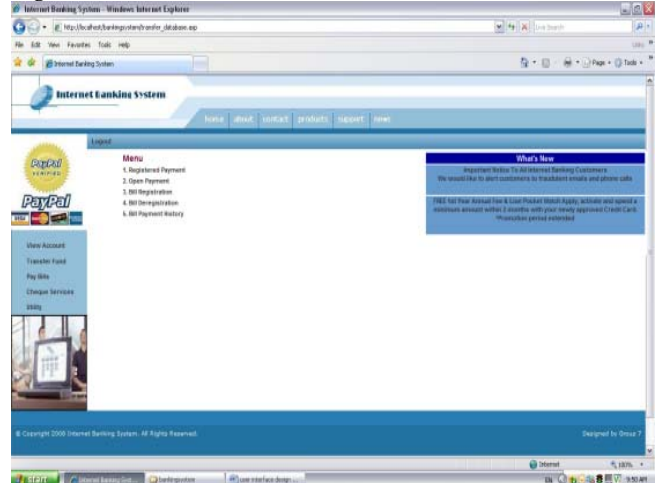

Fig 13. Pay Bills – Pay Bills Main Menu

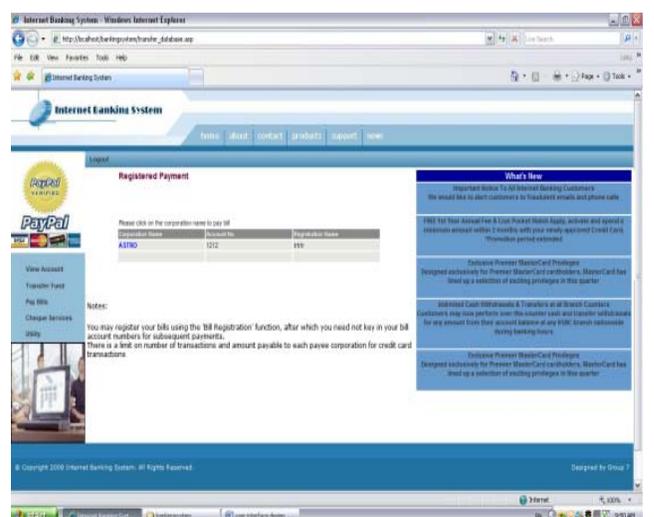

Fig 14. Pay Bills – Registered Payment Shows List of Registered Corporations

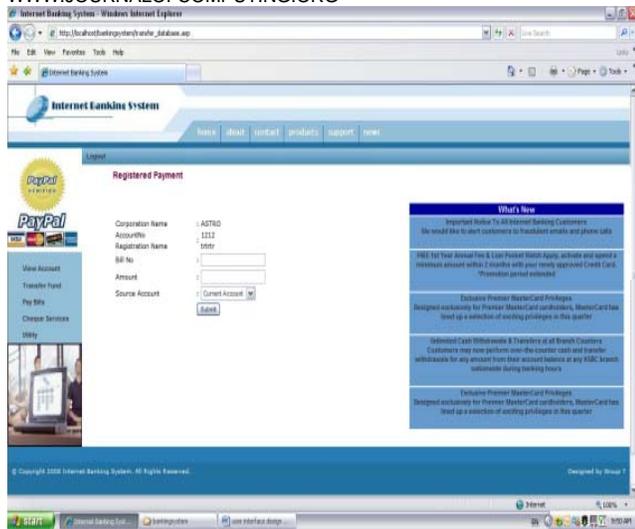
Fig 15. Pay Bills – Registered Payment Form for Selected Corporation

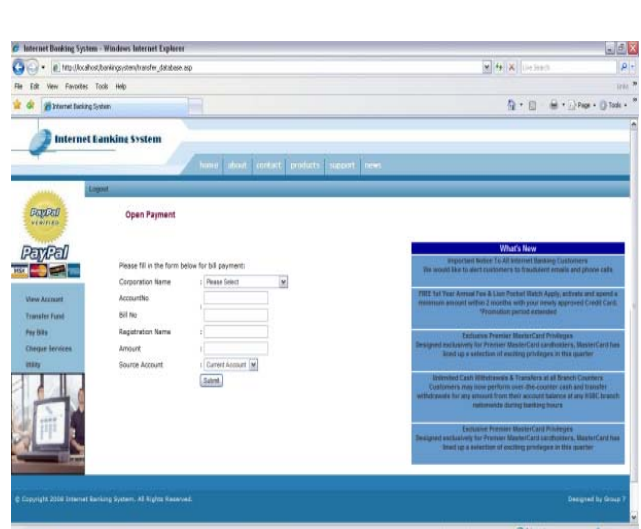
Fig 18. Pay Bills – Open Payment Form

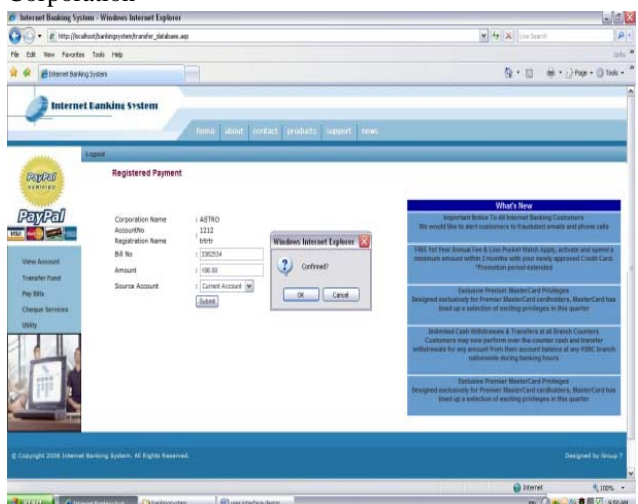
Fig 16. Pay Bills – Registered Payment Confirmation Pop Up Message

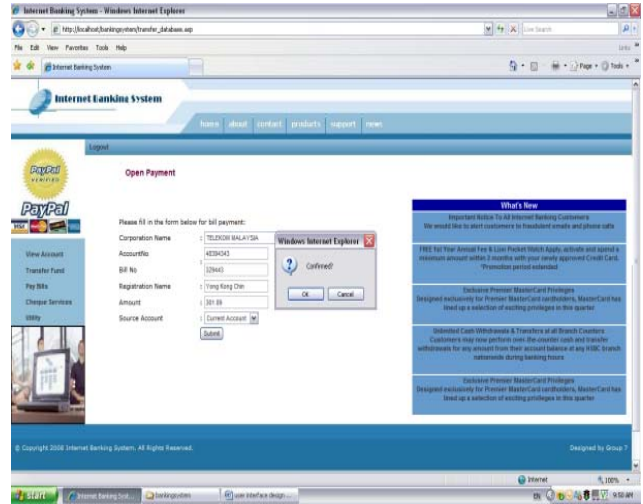
Fig 19. Pay Bills – Open Payment Confirmation Pop Up Message

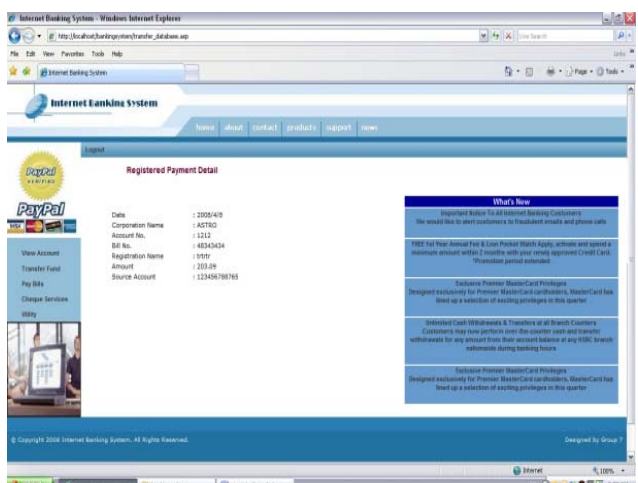
Fig 17. Pay Bills – Registered Payment Successful

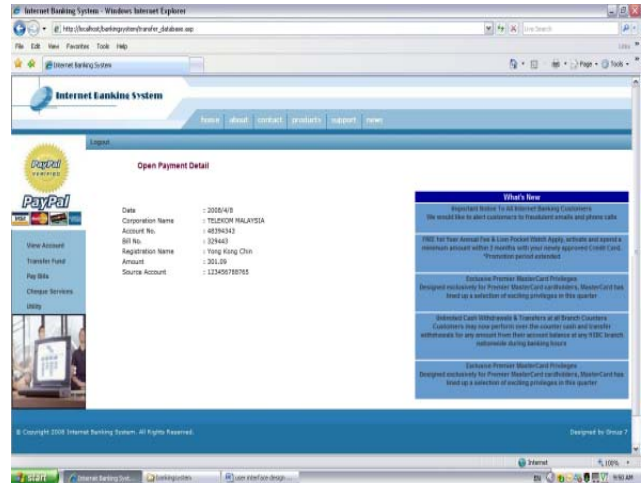
Fig 20. Pay Bills – Open Payment Successful

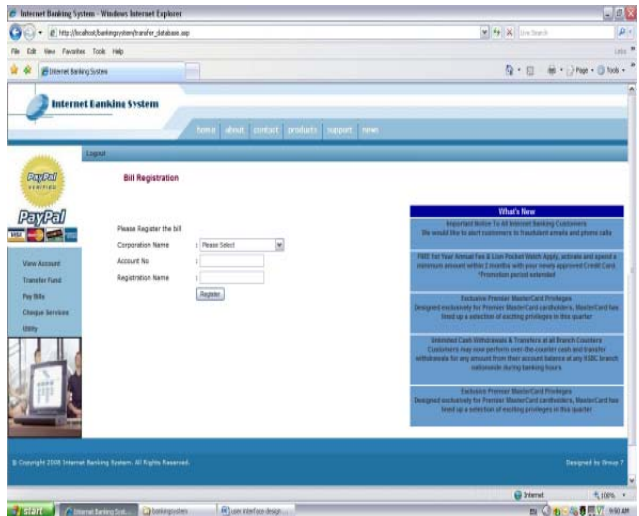
Fig 21. Pay Bills – Bill Registration

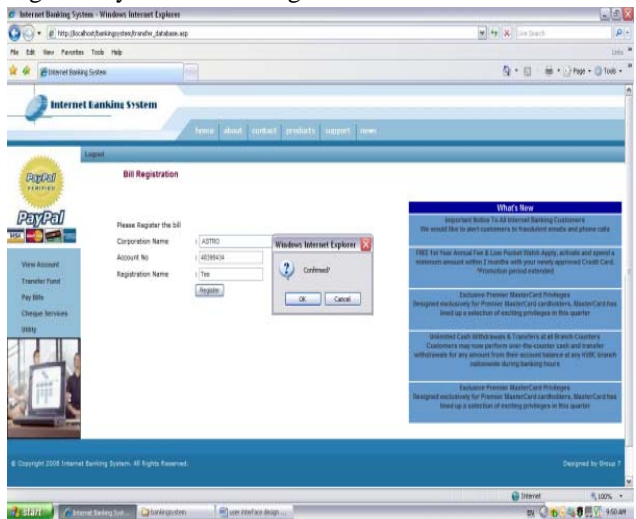
Fig 22. Pay Bills – Bill Registration Confirmation Message

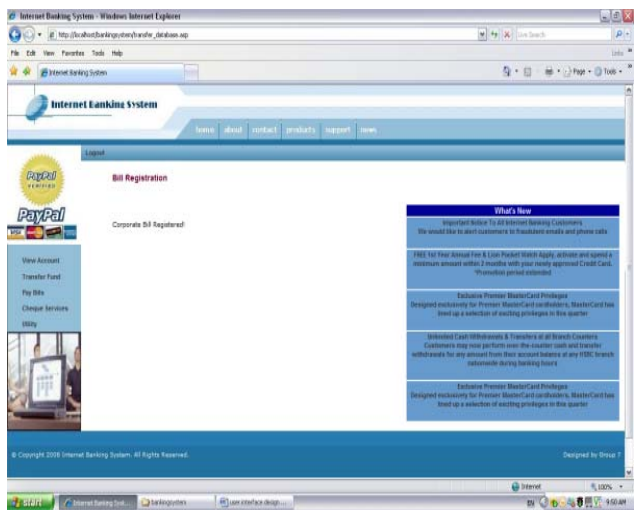
Fig 23. Pay Bills – Bill Registration Successful

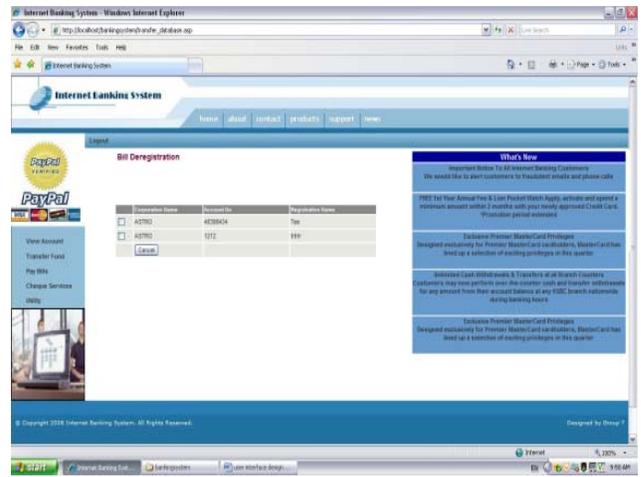
Fig 24. Pay Bills – Bill Deregistration Lists of corporations

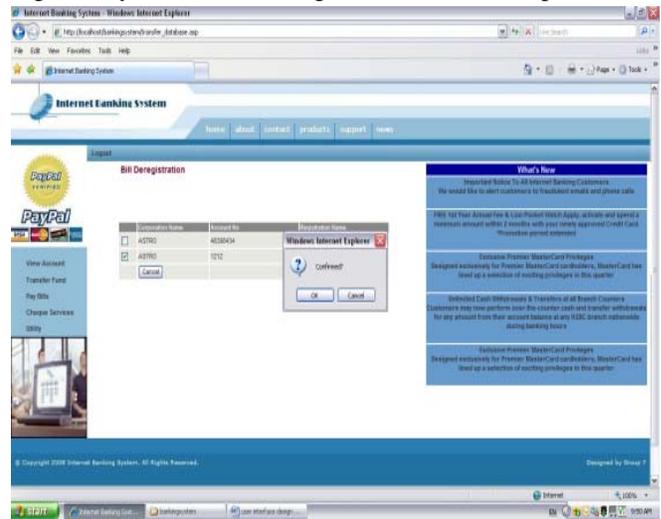
Fig 25. Pay Bills – Bill Deregistration Confirmation Messag

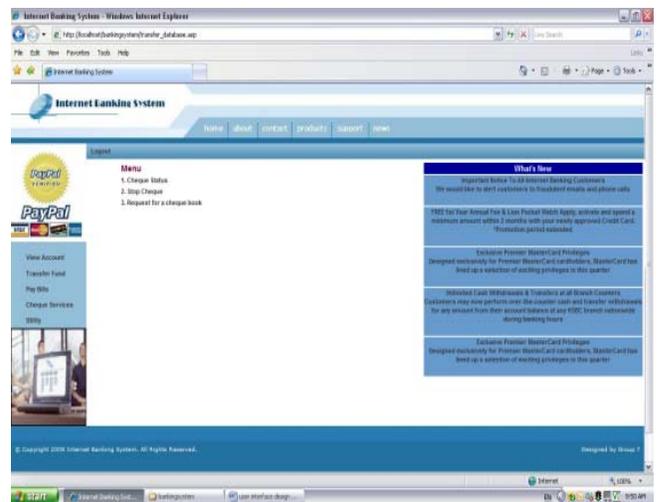
Fig 26. Cheque Services – Cheque Services Main Menu

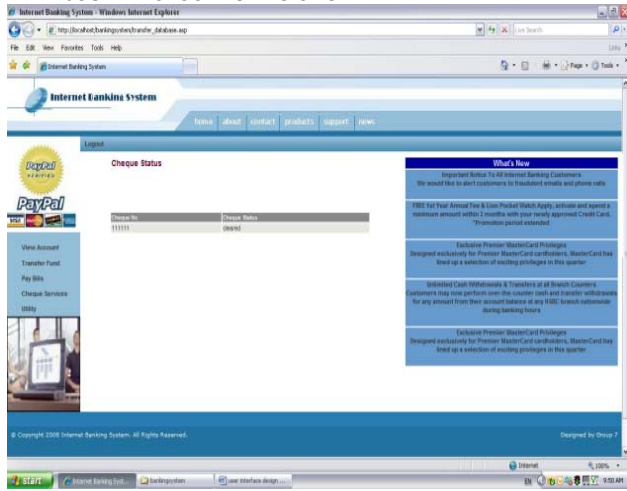
Fig 27. Cheque Services – View Cheque Status

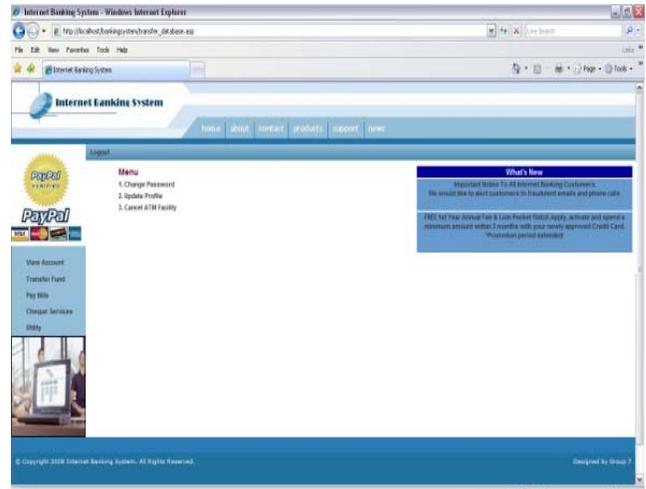
Fig 30. Utility – Utility Main Menu

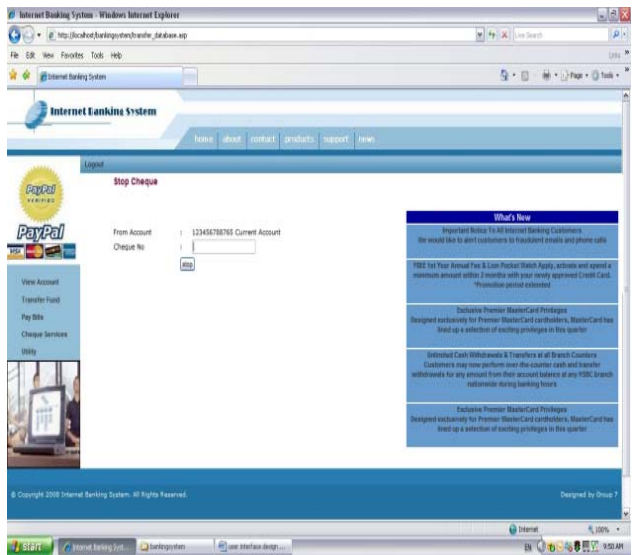
Fig 28. Cheque Services – Stop Cheque Form

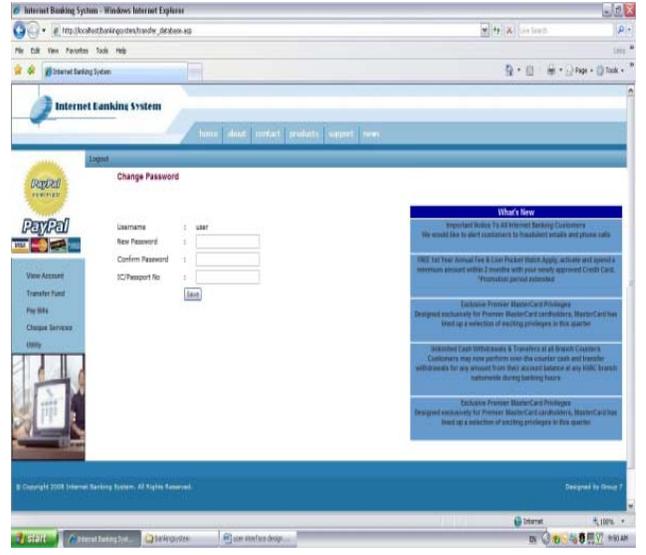
Fig 31. Utility – Change Password Form

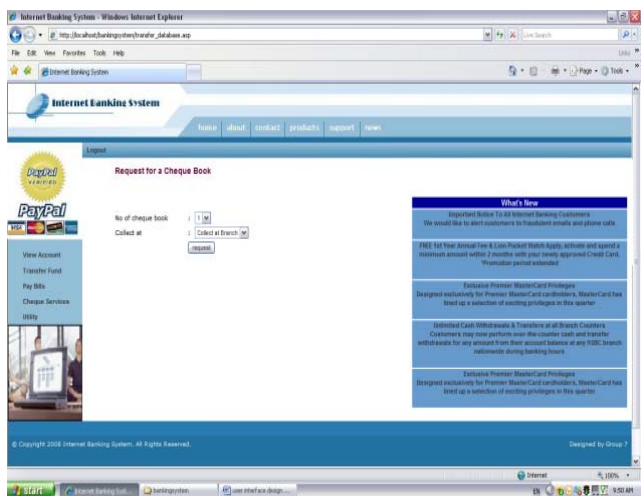
Fig 29. Cheque Services – Request for Cheque Book

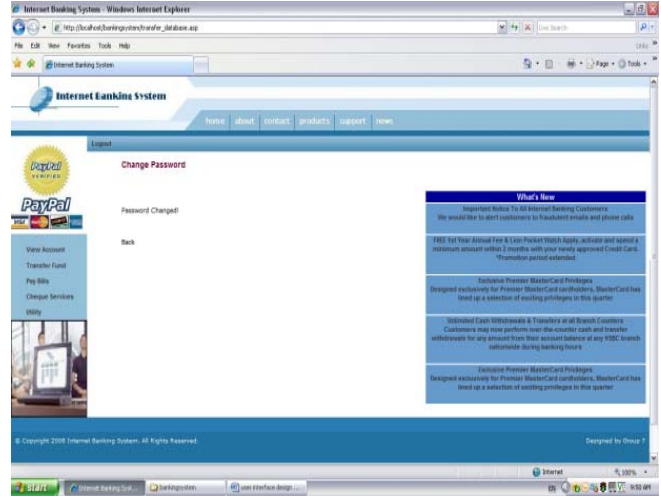
Figure 32: Utility – Password Successfully Changed

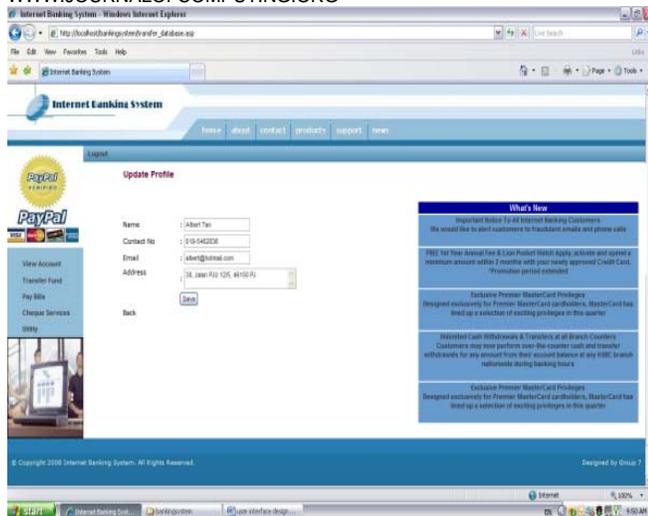

Figure 33: Utility – Update Profile Form

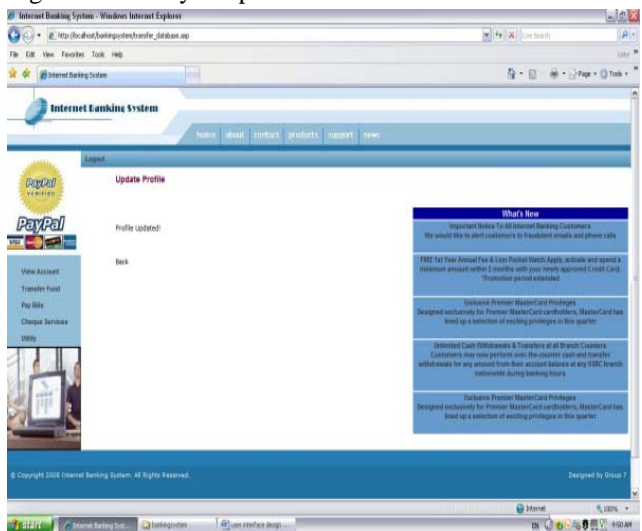

Figure 34: Utility – Profile Updated

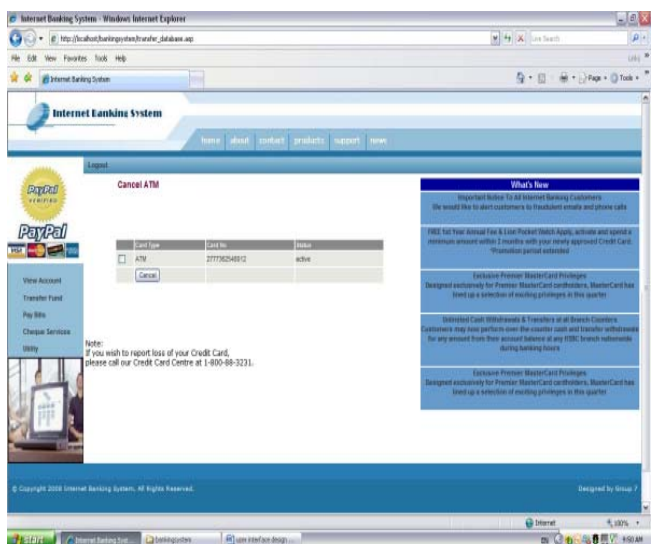

Figure 35: Utility – Cancel ATM Card

## 3. SYSTEM EVALUATION & TESTING

Software requirements specification should be a comprehensive statement of a software system's intended behavior. Unfortunately, requirements specifications are often incomplete, inconsistent, and ambiguous. We know that many serious conceptual errors are introduced in this first stage of software development-errors introduced during the requirements stage have been shown to be more difficult and more expensive to correct than errors introduced later in the lifecycle, and they are more likely than implementation errors to be safety critical .Therefore, it is important to provide methods and techniques to eliminate requirements-related errors as early as possible. To provide analysis procedures to find errors in specifications, it is first necessary to determine the desirable properties of a Specification.

The completeness for this banking system shown that all services required are defined in the system as the following.

- **Login Page**
  - ✓ Login Page Main Menu

Refer to Figure 1
  - ✓ Invalid User Page

Refer to Figure 2
  - ✓ Valid User Page

Refer to Figure 3

- **View Account**
  - ✓ View Account Main Menu (Account Type)

Refer to Figure 4
  - ✓ Transaction History – Current Account

Refer to Figure 5
  - ✓ Transaction History – Saving Account

Refer to Figure 6

- **Transfer Fund**
  - ✓ Transfer Fund Main Menu

Refer to Figure 7
  - ✓ Transfer Funds Form

Refer to Figure 8
  - ✓ Transfer Fund Confirmation Pop Up Message

Refer to Figure 9
- ✓ Transaction Successful

Refer to Figure 10
- ✓ Transaction History

Refer to Figure 11
- ✓ Transfer Funds Detail

Refer to Figure 12

- ➢ **Pay Bills**
  - ✓ Pay Bills Main Manu

Refer to Figure 13
- ✓ Registered Payment Shows List of Registered Corporation

Refer to Figure 14
- ✓ Registered Payment Show Form for Selected Corporation

Refer to Figure 15
- ✓ Registered Payment Confirmation Pop Up Message

Refer to Figure 16
- ✓ Registered Payment Successful

Refer to Figure 17
- ✓ Open Payment

Refer to Figure 18
- ✓ Open Payment Confirmation Pop Up Message

Refer to Figure 19
- ✓ Open Payment Successful

Refer to Figure 20
- ✓ Bill Registration

Refer to Figure 21
- ✓ Bill Registration Confirmation Message

Refer to Figure 22
- ✓ Bill Registration Successful

Refer to Figure 23
- ✓ Bill Deregistration Lists of corporations

Refer to Figure 24
- ✓ Bill Deregistration Confirmation Message

Refer to Figure 25
- ✓ Bill Payment History

Refer to Figure 26

- ➢ **Cheque Services**
  - ✓ Cheque Services Main Manu

Refer to Figure 27
- ✓ Cheque Status Form

Refer to Figure 28
- ✓ View Cheque Status

Refer to Figure 29
- ✓ Stop Cheque

Refer to Figure 30
- ✓ Request for cheque book

Refer to Figure 31

- ➢ **Utility**
  - ✓ Utility

Refer to Figure 32
- ✓ Change Password

Refer to Figure 33
- ✓ Change Password Confirmation Pop Up Message

Refer to Figure 34
- ✓ Password Successfully Changed

Refer to Figure 35
- ✓ Update Profile

Refer to Figure 36
- ✓ Update Profile Confirmation Pop Up Messages

Refer to Figure 37
- ✓ Profile Updated

Refer to Figure 38
- ✓ Cancel ATM Card

Refer to Figure 39

## 4. CONCLUSION

The internet banking system is aiming to provide users with the easiest way to access to their bank account and do some banking transactions anytime at anywhere without the need to go to the bank. We have basically come out with all functional and non-functional requirements for this internet banking system to make it a success. The functional requirements are login, logout, view account, transfer funds, pay bills, cheque services and utility. The basic functions of internet banking system web portal will be built from all these requirements. However, to further make the internet banking system web portal a successful one, non-functional requirements such as safety, security, performance and quality attributes will be required.

## ACKNOWLEDGMENT

This research was fully supported by "King Saud University", Riyadh, Saudi Arabia. The author would like to acknowledge all workers involved in this project that had given their support in many ways, aslo he would like to thank in advance Dr. Musaed AL-Jrrah, Dr. Abdullah Alsbail, Dr. Abdullah Alsbait. Dr.Khalid Alhazmi , Dr. Ali Abdullah Al-Afnan, Dr.Ibrahim Al-Dubaian and all the staff in king Saud University especially in Applied Medical Science In "Al-Majmah" for thier unlimited support, without thier notes and suggestions this research would not be appear.

**Rami Alnaqeib-** he is master student in the Department of Information Technology / Faculty of Computer Science and Information Technology/University of Malaya / Kuala Lumpur/Malaysia, He has contribution for many papers at international conferences and journals

**Hamdan Al-Anazi**: He has obtained his bachelor degree from "King Saud University", Riyadh, Saudi Arabia. He worked as a lecturer at Health College in the Ministry of Health in Saudi Arabia, and then he worked as a lecturer at King Saud University in the computer department. Currently he is Master candidate at faculty of Computer Science & Information Technology at University of Malaya in Kuala Lumpur, Malaysia. His research interest on Information Security, cryptography, steganography, Medical Applications, and digital watermarking, He has contributed to many papers some of them still under reviewer.

**Dr.Hamid. A.Jalab**: Received his B.Sc degree from University of Technology, Baghdad, Iraq. MSc & Ph.D degrees from Odessa Polytechnic National State University 1987 and 1991, respectively. Presently, Visiting Senior Lecturer of Computer System and Technology, Faculty of Computer Science and Information Technology, University of Malaya, Malaysia. various international and national conferences and journals, His interest area are Information security (Steganography and Digital watermarking), Network Security (Encryption Methods) , Image Processing (Skin Detector), Pattern Recognition , Machine Learning (Neural Network, Fuzzy Logic and Bayesian) Methods and Text Mining and Video Mining.

**Mussab alaa Zaidan -** he is master student in the Department of Information Technology / Faculty of Computer Science and Information Technology / University of Malaya/ Department /Kuala Lumpur/Malaysia, He has contribution for many papers at international conferences and journals.

**Ali K.Hmood -** he is master student in the Department of Software Engineering / Faculty of Computer Science and Information Technology/University of Malaya /Kuala Lumpur/Malaysia, He has contribution for many papers at international conferences and journals.

.